# Revisiting the Core Ontology and Problem in Requirements Engineering[*]

Ivan J. Jureta
PReCISE
University of Namur
iju@info.fundp.ac.be

John Mylopoulos
Dept. of Computer Science
University of Toronto
jm@cs.toronto.edu

Stéphane Faulkner
PReCISE
University of Namur
sfaulkne@fundp.ac.be

October 22, 2018

## Abstract

In their seminal paper in the ACM Transactions on Software Engineering and Methodology, Zave and Jackson established a core ontology for Requirements Engineering (RE) and used it to formulate the "requirements problem", thereby defining what it means to successfully complete RE. Given that stakeholders of the system-to-be communicate the information needed to perform RE, we show that Zave and Jackson's ontology is incomplete. It does not cover all types of basic concerns that the stakeholders communicate. These include beliefs, desires, intentions, and attitudes. In response, we propose a core ontology that covers these concerns and is grounded in sound conceptual foundations resting on a foundational ontology. The new core ontology for RE leads to a new formulation of the requirements problem that extends Zave and Jackson's formulation. We thereby establish new standards for what minimum information should be represented in RE languages and new criteria for determining whether RE has been successfully completed.

---

# 1  Introduction

A decade ago Zave and Jackson [21] observed that the field of Requirements Engineering (RE) left behind simplistic approaches to understanding what a system will do in favor of novel and varied terminology, methods, languages, tools, and issues considered to be critical. They went on to define a core ontology that establishes minimum standards for what information should be represented in any RE language. This allowed them to formulate the "requirements problem", which determines exactly what it means for RE to be successfully completed. Various representations and interpretations of the core ontology have since been suggested and used. However, there is limited consensus on the precise meaning of the core ontology in RE. Rather, a requirement is variously understood as (describing) a purpose, a need, a goal, a functionality, a constraint, a quality, a behavior, a service, a condition, or a capability. Terms such as "nonfunctional requirement", "softgoal", "preference", "priority" only add to the confusion. It is difficult to compare or combine contributions, and identify conceptual overlaps.

*The only relevant core ontology is one which helps the software engineer in solving the requirements problem.* Zave and Jackson's [21] provided an elegant characterization of the requirements problem by relying on three concepts that constitute their core ontology. A "requirement" is an optative (i.e., desired) property of the environment, which includes the system-to-be and its relevant surroundings. A "domain assumption" is an indicative property, describing the environment as it is and in spite of the system-to-be. An optative property, intended to be directly implementable and support the satisfaction of requirements, is a fragment of a "specification". From there on, Zave and Jackson suggest that *the requirements problem amounts to finding the specification* $\mathbf{S}$ *that for given domain assumptions* $\mathbf{K}$ *satisfies the given requirements* $\mathbf{R}$. If all three are written in a mathematical logic, the problem is solved once the engineer finds $\mathbf{S}$ such that $\mathbf{K}, \mathbf{S} \vdash \mathbf{R}$.

This characterization is, unfortunately, deficient. Firstly, it does not allow for partial fulfilment of (some) requirements, e.g., non-functional ones. This is in sharp contrast to RE practice where it is not even clear what it means to fully satisfy a non-functional requirement. Secondly, the characterization leaves no room for one specification to be "better" than another, given domain assumptions K and requirements R. Thirdly, important notions such as those of non-functional requirement and preference (or, "nice to have" requirements) are left out of the framework, with no substitutes to compensate for this omission.

Our aim is to suggest a core ontology that spans much of ongoing research in RE and is grounded in the communication between the stakeholders and the engineer (§2–3). This allows us to provide a new formulation of the requirements problem that responds to the above limitations (§4). We then draw conclusions, relate to comparable discussions, and highlight important directions for future work (§5).

## 2 Baseline

To select and define the concepts in our core ontology, we start from the simple premise that *stakeholders communicate information to the software engineer* in the RE process. Stakeholders include people, but also, e.g., legacy systems for which the documentation is available. The engineer's task is to classify information as requirements or otherwise (e.g., domain assumptions), then use it in writing a specification. While this appears to be an obvious understanding of the overall RE process, it has important consequences on the core ontology, and is the key source of novelty with regards to available research. *Utterances that stakeholders make in communicating with the engineer are actions intended to advance stakeholders' personal desires, intentions, beliefs, and attitudes*, in the aim of ensuring that the engineer can produce a specification that then leads to the system responsive to the communicated concerns. It is important to observe here that, *if we know the various kinds of communicative actions at disposal of the stakeholders, we can delimit the scope of the core ontology to concepts that allow us to represent all of the communicated concerns.* By looking at the very origin of all requirements and related information, that is, the communication process, our concepts are instead grounded there: they cover the communicative actions available to the stakeholders.

We turn to speech-act theory to better understand the communication process. There, communication is considered as action [17]: a speaker makes an utterance in an attempt to change the state of the world. What distinguishes speech acts from non-speech actions is the domain of the speech act, that is the part of the world that the speaker wishes to modify through the act: in speech acts, this is mostly the mental state of the hearer. In our setting, we have stakeholders as speakers, the engineer as the hearer. Given an utterance, *the engineer should distinguish its content from its illocutionary force in order to know to represent the content as a requirement or otherwise.* For example, when I say "Book the plane ticket" (or if I wrote in some documentation given to the engineer), I may be expressing a desire (illocutionary force) that a plane ticket be booked (content, i.e., what I desire). Being desired, the content indicates a requirement that I hope the system will ultimately satisfy.

*Depending on the illocutionary force, the engineer will classify the content of the communication as requirements or otherwise, and thereby differently represent and act upon the communicated content: the core ontology should include concepts that cover all kinds of illocutionary force.* Depending on illocutionary force, Searle distinguishes the following kinds of speech acts [18]: (i) *assertives*, which assert a proposition that the speaker believes is true; (ii) *directives*, which convey a proposition that the speaker wants to see become true; (iii) *commissives* stating what the speaker intends to (do to) make a proposition true; (iv) *expressives* that convey a speaker's emotion/attitude about herself or the hearer; (v) *declarations* which by the very act of being stated make a proposition true; and (vi) *representative declaratives*, which recognize the truth of a proposition that has been made true by a declaration.

Any core ontology for RE should feature a minimal set of concepts and relationships that together can capture and distinguish between beliefs, desires, intentions, and attitudes (as they are communicated by expressives).

Once we know what to cover with the core ontology, we need to produce definitions of the concepts. To ensure an acceptable degree of rigor, we define our concepts by mapping them to concepts defined in a restricted vocabulary, namely a *foundational* ontology for which the underlying philosophical choices are clear and ontological categories well-delimited and openly discussed. This makes our assumptions clearer and easier to discuss in future efforts.

A foundational ontology is a theory about the abstract domain-independent categories in the real world. Its main purposes are to act as a starting point for building new ontologies and as a reference point for easy and rigorous comparisons among ontological approaches. Among the various available foundational ontologies (for comparisons, see [14, 4]), we choose DOLCE [14]. DOLCE rests on intuitively attractive ontological choices for the present discussion. In particular, DOLCE is descriptive in that it aims to capture the ontological categories underlying natural language and human common sense. Moreover, categories in DOLCE are defined to reflect notions that depend on human perception, cultural imprints, and social conventions. Decisions taken over the course of RE are informed by stakeholders having perceptual bias, individual cultural and other backgrounds, and specific social conventions. Not recognizing such bias could hardly be an appropriate choice at present: the resulting RE ontology would be removed from the reality of the problem that we intend to resolve. Categories in the ontology are therefore conceived as cognitive artifacts ultimately depending on human perception, cultural imprints, and social conventions. DOLCE is an ontology of particulars that distinguishes four basic categories – any particular is either an *endurant*, a *perdurant*, a *quality*, or an *abstract*. All proper parts of an endurant are wholly present at any time, whereas a perdurant accumulates parts over time (only some of its parts are present at any time). A quality is a basic perceivable and measurable entity that inheres to other entities (e.g., the color of this desk). It is DOLCE's concept of *quality* that we use extensively below in our core ontology for RE. In DOLCE, qualities are basic entities that one perceives and measures. A quality differs from the concept of "property" in that the former is a particular, while the latter is a universal. Every entity comes with some qualities. Also, qualities can inhere in other qualities. Qualities belong to a finite set of quality types, such as, e.g., size, color, etc. and are characteristic for – or inhere in – specific individuals. A quality (e.g., color) is distinguished from its "value" (e.g., some shade of red); the latter is called a "quale" and describes the position of an individual in a certain conceptual space, that is "quality space." Two particulars having exactly the same color thus carry color qualities which have the same position in the color space, or, in other words, have the same color quale. In terms of basic ontological categories in DOLCE, communication is a perdurant, and speech acts are sub-categories of communication. Mental states expressed by speech acts are endurants. More precisely, they are sub-categories of DOLCE's notion of mental object. In the remainder, we use DOLCE with explicit speech acts and mental states as the foundational ontology that gives us a restricted vocabulary for defining the core ontology for RE.

## 3   Core Ontology for RE

This section introduces the core ontology and explains the rationale behind it. Facing a speech act, the engineer should first distinguish its type (i.e., assertive, commissive, etc.), then separate its modus (modality) from its dictum (content). We have four modalities that correspond to the mental states underlying the speech acts: belief (B), desire (D), intention (I), and attitude (A). The engineer associates a modality to the content of a given speech act, and then proceeds to determine if the obtained result is an instance of a concept or relationship in our core ontology. The concepts of our core ontology are *goal*, *softgoal*, *quality constraint*, *plan*, and *domain assumption*; and the relationships are *attitude*-based optionality and preference, *justified*

*approximation*, and *non-monotonic consequence*.

Below, we first explain the association of modalities to communicated content on the basis of speech act type (§3.1). We subsequently discuss each concept of the core ontology. We indicate throughout the section precisely how our ontology departs from Zave and Jackson's.

## 3.1 Classifying Communicated Content

Consider the problem of designing a system for booking flights online, for which a stakeholder suggests:

(Ex.1) *No business seat has a lower price than an economy seat.*

If the stakeholder believes the above is already true, the statement is an assertive speech act. If the stakeholder intends to institute the above as a rule, the statement is a declarative speech act. If the stakeholder merely reiterates that the above applies, it is a representative declarative.

(Ex.2) *I hope/wish/desire/expect booking to be always confirmed with the new system.*

(Ex.3) *I will ensure that booking is always confirmed.*

(Ex.4) *It is preferred that the booking confirmation be sent quickly after booking.*

Statement Ex.2 is directive as it indicates what is desired. Ex.3 is commissive for it points out the stakeholders' intention on bringing about something desired. Finally, Ex.4 is expressive, as it makes explicit stakeholder's attitude on how the booking confirmation is to be sent to the user. While the illocutionary point is not apparent from written text (as in the examples above), it is not absent [19]: when given documentation, the engineer must determine if it expresses desires, facts, or otherwise.

Following Zave and Jackson [21], we know that the critical distinction between a requirement and a domain assumption lies in that the former expresses something desired, while the latter something that is already the case, or is/will be the case regardless of the system. It is not difficult to see that Ex.1 is thereby a domain assumption, while Ex.2 is a requirement. We also know from Zave and Jackson that a specification together with domain assumptions can satisfy requirements. Moreover, we know from Cohen and Levesque [5] that an agent adopts intention X if (i) it believes X is possible, (ii) it does not believe it will not bring about X, (iii) it believes it will bring about X, (iv) it does not intend all the side effects of bringing about X, and (v) it invests effort in trying to bring about X. *Specification thereby involves intentions*: agents (including, e.g., stakeholders and the system-to-be) choose how to act and commit to do so, and this in order to bring about states of the world in which requirements are satisfied and domain assumptions are not violated. Hence, while desires lead to requirements and beliefs to domain assumptions, intentions lead to specifications. Ex.3 is part of a specification.

We see therefore that Zave and Jackson's domain assumption, requirement, and specification concepts cover propositional attitudes (i.e., belief, desire, and intention) as they are usually conceptualized in philosophy [16]. More can be, and often is communicated than propositional attitudes. Looking at the notion of attitude in psychology [1] (which is what expressive speech acts convey), attitude is identified most closely with affect, i.e., a general evaluative reaction (e.g., "I like Y" or "I like Y more than Z"). Ex.4 is an expressive as it communicates an attitude. Zave and Jackson's core ontology is incomplete in that attitudes are missing, which, as we show further on, has significant consequences on the formulation of the requirements problem.

Our approach consists first of classifying communicated content by associating modalities to it, whereby the choice of modality depends on the kind of speech act used in communication. Rules for associating modalities to content are:

- The content $\phi$ of an assertive or declarative or representative declarative speech act that stakeholders communicated to the software engineer is a belief, $\mathsf{B}\phi$.

- The content $\phi$ of a directive speech act that stakeholders communicated to the software engineer is a desire, $\mathsf{D}\phi$.

- The content $\phi$ of a commissive speech act that stakeholders communicated to the software engineer is an intention, $\mathsf{I}\phi$.

- The content $\phi$ of an expressive speech act that stakeholders communicated to the software engineer is an attitude, $\mathsf{A}\phi$.

If we encounter conjunctive, disjunctive, or conditional (e.g., if [atomic speech act 1] then [atomic speech act 2]) linking of speech acts, atomic speech acts are separately considered when associating modalities to their content. Observe the parallels between the above and Zave and Jackson's ontology: their requirement here corresponds to $\mathsf{D}\phi$, while $\mathsf{I}\phi$ and $\mathsf{B}\phi$ correspond to, respectively, a specification (that is, a fragment thereof) and a domain assumption. There is no departure in this respect from the accepted intuitions in RE: requirements cover what is desired, while domain assumptions concern what is true. Intentions give specification fragments for they determine what will be done to satisfy requirements.

### 3.2 Domain Assumption

Beliefs communicated by way of assertive, declarative, or representative declarative speech acts constrain the possible states of the world only to those in which beliefs are not violated. They correspond to domain assumptions, which should not be violated by the sytem or its relevant surroundings, as the following definition indicates.

**Definition 1.** *Believed content, i.e., $\phi$ in $B\phi$, communicated by way of assertive, declarative, or representative declarative speech acts is a **domain assumption**, denoted generically $k$.*

Believed content is thus considered a member of the set **K** of domain assumptions that the software engineer elicits and should account for when building a specification. **K** therefore contains the content of the kinds of speech acts indicated in Definition 1.

(Ex.5)  *Standard credit card symbols must be displayed whenever the customer is asked to make a payment.*

If the statement above is the content of an assertive speech act, the engineer sees it as a domain assumption. In other words, the content above delimits possible states of the world only to those in which payment card symbols are displayed to the customer, whenever she is asked to make a payment. Consequently, a specification is not acceptable if it leads to states in which the above is violated.

### 3.3 Goal, Quality Constraint, Softgoal

The concepts of goal and quality constraint are introduced to cover the classical taxonomic dimension for the requirement concept: the distinction between the notions of functional and nonfunctional requirement. It is widely accepted in RE that functional requirements refer to what the system does, as opposed to how well the system does what it does. The latter is covered by nonfunctional requirements. Ex.6 thus gives a functional requirement.

(Ex.6)  *Once the payment for the flight is confirmed, book the flight.*

(Ex.7)  *It should be possible to fully book a flight through less than 5 different screens.*

Ex.7 indicates how well booking performs for it identifies a measure on the behavior of the system (i.e., the number of different screens the user needs to go through to book a flight). In other words, the second statement characterizes flight booking, since it points to the existence of *qualities* for which only some (sets of) possible *values* are desired. It follows that the functional vs. nonfunctional distinction is grounded in the notion of quality in the sense of DOLCE. Namely, a requirement that describes qualities and constrains quality values is a nonfunctional requirement, and that the requirement is otherwise a functional requirement. This still remains within accepted intuitions in RE: any functional requirement will describe what the system does, whereas a nonfunctional requirement will refer to measures on the system's doings, thus allowing us to characterize how well the system behaves. Although uncontroversial, this separation obtains a solid foundation herein, leading to the following definitions.

**Definition 2.** *Desired content, i.e., $\phi$ in D$\phi$, communicated by way of a directive speech act is a **quality constraint**, denoted **q**, if and only if $\phi$ describes qualities and constrains quality values. Described qualities must have quality space with a well-defined and shared structure.*

**Definition 3.** *Desired content, i.e., $\phi$ in D$\phi$, communicated by way of a directive speech act is a **goal**, denoted **g**, if and only if $\phi$ neither describes qualities nor constrains quality values.*

It should be possible to verify whether a system satisfies requirements before delivering it to the stakeholders. Any goal and any quality constraint is verifiable, in the sense that the software engineer can check at any time whether the system satisfies the chosen goal or quality constraint. Goals clearly are verifiable for the software engineer can determine whether what is functionally required is indeed delivered by the system. Whether the flight is booked whenever the payment is confirmed is not a matter of cognitive bias or ill-defined verification criteria: the flight is either booked or not. It is the second sentence in Definition 2 that ensures verifiability for quality constraints: a quality constraint requires a well-defined quality space (e.g., possible values are known, values and their relationships are precisely defined, and so on) *and* the quality space must be shared among the stakeholders.

But nonfunctional requirements are taken to include abstract considerations such as security, safety, maintainability, convenience, and so on. The question then is: How do these abstract notions relate to our quality constraint concept? In other words, does desiring, e.g., "high convenience in flight booking" give us a quality constraint or something else?

Given that a quality (in our foundational ontology of choice) is a perceivable and measurable entity that inheres in other entities, there must be a quality space and values for, e.g., convenience in order to call it a quality, and thereby state that asking for, e.g., "high convenience" is a quality constraint. Now, we know that people have the ability to evaluate convenience since we observe that they can say to what extent a system is convenient. We also know that such evaluations are not necessarily (or rarely are) shared: while one person gives a strong favorable evaluation, another one may disagree. It is reasonable to argue that any evaluation requires that a somehow structured quality space exists. We could consequently conclude that high convenience is a quality constraint. To do so is, however, a mistake for the following reasons:

- Divergent evaluations of the same phenomenon mean that the structure of the underlying quality space is dependent on each individual's cognitive bias (i.e., the structure is subjective).

- While there seems to be some kind of quality space at each individual, people are rarely (if ever) capable of defining and/or conveying the structure of that quality space (e.g., for convenience) in a precise manner, so that verifiability remains elusive.

- Subjective structure of the quality space counters the need for a quality space with a structure shared among the stakeholders, which would subsequently facilitate verification.

To aim the verifiability of quality constraints, we require well-defined quality spaces (e.g., as in metrics) and we need quality spaces shared among the stakeholders, or such that the stakeholders can accept to share these in relation to the system of interest. Qualities can take the form of measures for which we make explicit the desired values, that is, we define quality constraints. Measures will have a particular level of measurement: we can have nominal, ordinal, interval, ratio or other level of measurement [13]. There is no restriction, e.g., on having necessarily measures that have rich sets of permissible transformations. We conclude thus that while quality constraints cover some of nonfunctional requirements, abstract nonfunctional requirements on, e.g., convenience, security, maintainability, and so on, are *not* quality constraints: we call them *softgoals* instead.

**Definition 4.** *Desired content, i.e., $\phi$ in $\mathsf{D}\phi$, communicated by way of a directive speech act is a **softgoal**, denoted $\hat{q}$, if and only if $\phi$ describes qual-*

*ities or constrains quality values, whereby the described qualities must have a quality space with a subjective and/or ill-defined structure.*

The following statement, communicated by way of a directive speech act gives us a softgoal.

(Ex.8) *Flight booking should be convenient.*

A salient characteristic of softgoals is that they cannot be satisfied to the ideal extent, not only because of subjectivity, but also because the ideal level of satisfaction is beyond the resources available to (and including) the system. It is therefore said that a softgoal is not satisfied, but *satisficed*. The software engineer seeks the specification that satisfies to the highest extent compared to the *considered* alternatives. This contrasts to seeking (under resource constraints) the best among all possible alternatives, which amounts to optimization.

But how then does one cope with softgoals? Mylopoulos and colleagues [15] introduced so-called contribution links between goals and softgoals to indicate that bringing about states in which the goal is satisfied contributes positively or negatively to the satisficing of the softgoal. The aim with contribution links is to enable the engineer to claim that satisfying goals in some particular way leads to some degree of satisfaction of the softgoals. While contribution links are attractive for they allow side-by-side comparison of alternative system structures (e.g., [2]), their semantics remain vague, which has led to various interpretations and use. To properly verify whether and to what extent satisficing occurs, we need to understand more precisely how goal, quality constraint, and softgoal are related. This is necessary in order to show later on (§4) the role of softgoals in the requirements problem.

When the engineer aims to ensure convenience of the flight booking process, she will try to understand how various values of measurable system behaviors affect stakeholders' perception of convenience. For instance, she may consider that 6 screens for flight booking correspond to a threshold level of convenience, and that convenience improves as the number of screens reduces, whereby anything less than 3 screens is unacceptable for the user is expected to fill out too big a form over too few screens. Doing this amounts to *approximating* a subjective and/or ill-defined quality space by one or more other, well-defined and shared quality spaces. A contribution link (in its usual interpretation – see, [15]) between quality constraints over the latter and the softgoals over the former thus indicates that the said quality constraints approximate the degrees of satisfaction of the given softgoals. Hence, there is correlation between the values of the quality spaces underlying the quality constraints, and the values in quality spaces underlying the softgoals, which leads us to the following definition for contribution links within our ontology.

There are two important nuances to understand at this point. First, classical contribution links in RE are labeled: e.g., if we choose a system design in which there will be less than 5 screens for flight booking, we will label the contribution link as positive (we would draw the symbol '+' on it). The label is not itself related to the sign of the correlation (e.g., the number of screens for flight booking is negatively correlated with the 'level' of convenience) and there is no absolute rule for deriving one from the other. Second, not all approximations (and thereby contributions and their labels) are equivalently appropriate or acceptable to the stakeholders. Evidently, the engineer cannot prove (in the formal sense) that it is indeed relevant to approximate convenience with the number of screens in flight booking and that the correlation is as described above; the closest feasible solution instead is to *justify* that a quality referred in a quality constraint approximates a quality referred in a softgoal. In other words, we are interested in justifying that a quality constraint approximates a softgoal, whereby the justification applies under two caveats: approximation stands *for all practical purposes within the given development project*, and approximation is justified only on grounds of the information *available* to the software engineer. New information may lead the engineer to revise the approximation that was previously justified: e.g., more details on how stakeholders evaluate convenience. The first caveat means that approximating convenience by (among others) the number of screens for booking is justified only locally, within the given project (it may be appropriate elsewhere, but that remains to be justified); the second caveat means that the approximation can become irrelevant, e.g., if the engineer finds out that stakeholders care in no way for the number of screens, and that convenience is merely related to the length of the form to fill when booking a flight.

**Definition 5.** *There is a **justified approximation**, denoted **jApprox**($\hat{q}$, $q$) if and only if there is a justification for the claim "$q$ approximates $\hat{q}$" and there is sufficient correlation between values in the quality space of $q$ and the quality space of $\hat{q}$.*

We speak of "sufficient" correlation, leaving the sufficient level to be fixed case by case, and for the given project. Moreover, we do not necessarily have precise correlation between a well- and an ill-defined quality space. Still, there should be evidence that it is reasonable to believe that correlation exists (e.g., based on the engineer's previous experience or otherwise).

What justification provides is a structured process by which we can determine whether there is enough support for the claim that **q** approximates **q̂**. For details, the reader is referred to, e.g., Simari and Loui's seminal work [20] (see also [3], and for recent discussions of justification in RE, see [10]). In conclusion to the issue mentioned earlier, to claim that softgoals are satisfied, the engineer must identify quality constraints that justifiably approximate these softgoals.

Compared to justified approximation, contribution links in RE do not recognize the importance of justification and correlation between values in the respective quality spaces. Classical contribution links are indirect when they link goals to softgoals. It is more appropriate to speak of correlation between measures over the system behaviors that satisfy goals (given by qualities referred in quality constraints) and softgoals, than to abstract from quality constraints, as is the case in classical contribution links. Our ontology highlights the importance of correlation and justification for the existence of a the approximation relationship, while not rejecting the classical contribution links. Justified approximation is one way to read contribution links, thus ensuring more rigor in the rationale that goes in their construction.

### 3.4 Plan

Stakeholders and the system-to-be commit to act in the aim of satisfying requirements and without violating the domain assumptions. Given the content of communicated intentions, the specification will delimit the ways in which the involved parties will bring about the desired and allowed states of the world. We use the term *plan* to denote the content of communicated intentions, and this regardless of granularity (i.e., we can have primitive and composite plans, but this is of no interest in the present discussion).

**Definition 6.** *Intended content, i.e., $\phi$ in $I\phi$, communicated by way of a commissive speech act is a **plan**, denoted **p**.*

Plans fit with the concepts introduced above as follows: the system and the stakeholders commit to execute plans in order to satisfy goals, whereby measures defined over plan executions act as qualities for which quality constraints can be defined. In the usual sense of documentation describing what the system does, a specification amounts to a definition of plans.

We are interested in plans that satisfy generic goals, such as "book a flight", "schedule a meeting" or "fulfill a book request", where the goal is instantiated at runtime for each particular flight to be booked, meeting to be scheduled, or book that is requested. Given a vector x of parameters, we write $\mathbf{r}(\mathsf{x})$ the goal that takes the given parameters when instantiated: e.g., x in "Book a flight" goal may involve departure and arrival dates and locations; when scheduling a meeting, the parameters would include the initiator, the time of the request, the list of participants, as well as the purpose of the meeting; for book requests, parameters would refer the author and title of the book, also the requestor and the time of the request.

Plans ought to be defined in such a way that their proper execution brings about states of the world in which goals verify and domain assumptions are not violated, all the while assuring that all quality constraints (including those that justifiably approximate softgoals) are satisfied. We can draw parallels here with Zave and Jackson's formulation of the requirements problem. For simplicity, let us leave out attitudes (we include them later - see §4) and recall that we would have $\mathbf{K}, \mathbf{S} \vdash \mathbf{R}$ according to Zave and Jackson [21]. Given the ontology we built up to now, it is tempting to suggest that the RE effort is successful once the engineer finds the specification $\mathbf{P}$ (that includes all relevant plans $\mathbf{p}$) such that $\mathbf{K}, \mathbf{P} \vdash \mathbf{G}, \mathbf{Q}$, whereby $\mathbf{Q}$ includes, for all $\hat{\mathbf{q}}$, one or more $\mathbf{q}$ standing in the justified approximation relationship to $\hat{\mathbf{q}}$ (i.e., for each $\hat{\mathbf{q}}$, we have $\mathbf{q}$ such that $\mathbf{jApprox}(\hat{\mathbf{q}}, \mathbf{q})$). $\mathbf{K}$ includes the domain assumptions $\mathbf{k}$, $\mathbf{P}$ carries the plans $\mathbf{p}$, and $\mathbf{G}$ incorporates the goals $\mathbf{g}$. The relation $\vdash$ is a monotonic one, so that (i) $\mathbf{K}, \mathbf{P}, \mathbf{G}, \mathbf{Q}$ must define a sound and complete theory, and (ii) we accept that if $\mathbf{K}, \mathbf{P} \vdash \mathbf{G}, \mathbf{Q}$, then, e.g., for $\mathbf{K} \subseteq \mathbf{K}'$, we still have $\mathbf{K}', \mathbf{P} \vdash \mathbf{G}, \mathbf{Q}$; in other words, monotonicity ensures that conclusions can never be "undone" by new information. Evidently, we can hardly ever guarantee to have a sound and/or complete theory for any but the smallest of toy systems, and we tend to continually encounter in practice new information that defeats previously valid conclusions (when, e.g., requirements are revised or domain assumptions change). Both $\mathbf{K}, \mathbf{S} \vdash \mathbf{R}$ and $\mathbf{K}, \mathbf{P} \vdash \mathbf{G}, \mathbf{Q}$ are thus removed from reality. Taking a non-monotonic consequence instead of the monotonic one resolves this problem, for it acknowledges that sound and complete theories are elusive for any realistic system, and that monotonicity does not apply. Taking, for example, the defeasible consequence relation $\vdash\!\sim$ (as in [20]) for non-monotonic satisfaction, we can suggest that the following conditions ought to hold for generic goals to be satisfied and domain assumptions to hold (for all allowed parameter values in all parameterization vectors of the form x): (1) there are quality constraints $\mathbf{q}$ in $\mathbf{Q}$ that justifiably approximate all softgoals $\hat{\mathbf{q}}$ in $\hat{\mathbf{Q}}$; and (2) $\mathbf{K}, \mathbf{P} \vdash\!\sim \mathbf{G}, \mathbf{Q}$.

If the above do hold, the software engineer can claim it justified to believe that all goals and quality constraints will be satisfied if the plans that the involved parties execute at runtime do not violate **P**. In other words, if the various parties succeed in their intentions, the desired and allowed states of the world will be reached and this while reaching values of qualities referred to in quality constraints. Having non-monotonicity instead of monotonicity indicates that there is no proof for this claim, merely that there is enough evidence for accepting the given claim, and this within the information available to the stakeholders and the software engineer and within the scope of the given development project. It is not necessary in practice for **P** to describe the actual plans that need to be taken: we encounter situations in which we know how precisely one of the parties ought to act, while the best we can do for other parties is to restrict potential ways in which they can act; e.g., we often treat third party web services as black boxes of functionality, so that our *plan* would be written as logical constraints on the inputs and outputs of the service.

### 3.5 Attitude

The content of expressive speech acts gives attitudes. Written down, an attitude amounts to a description of an evaluation in terms of degree of favor or disfavor [7]. Such degrees vary in sign (positive or negative) and in intensity, whereby the intensity of the valuation is relative: considering an object of attitude on its own involves implicit comparison to a set of objects perceived by the evaluator to be of the same kind [11]. As Kahneman and colleagues observe [11], "objects of attitudes [as the term is used in psychology] include anything that people can like or dislike, wish to protect or to harm, to acquire or to reject". Stakeholders can thus evaluate favorably or disfavorably, and with different intensity individual or alternative domain constraints (Ex.9), plans (Ex.10), goals (Ex.11), quality constraints (Ex.12), or softgoals (Ex.13).

(Ex.9) *Rule in Ex.1 is unacceptable for it limits the pricing options in case of promotions.*

(Ex.10) *Having the system confirm booking is preferred to having a person do it.*

(Ex.11) *It would be good if the user is informed of special flight offers.*

(Ex.12) *It is preferred to split the flight booking form over several screens than to show it fully one a single screen.*

(Ex.13) *Convenience of flight booking is more important than speed.*

As the software engineer hopes to construct high quality systems, she must be interested in attitudes: attitudes reflect stakeholders' level of satisfaction with plans, goals, domain constraints, and so on, thereby guiding the engineer during the decision-making on whether and how to define plans to satisfy goals, softgoals, quality constraints, and avoid violating domain constraints. What the engineer must have in order to rate alternative specifications is therefore an explicit account of stakeholders' attitudes.

**Definition 7.** *Attitudinal content communicated by way of an expressive speech act, i.e., $\phi$ in A$\phi$, is an **attitude**, denoted **a**, if and only if it evaluates in terms of favor or disfavor one or more elements constituting $\bm{K}$, $\bm{P}$, $\bm{G}$, $\bm{Q}$, or $\bm{\hat{Q}}$.*

It is important to understand how the introduction of attitude changes our ontology defined up to this point. By being an evaluation over components of $\bm{K}$, $\bm{P}$, $\bm{G}$, $\bm{Q}$, or $\bm{\hat{Q}}$, any attitude **a** is not a concept *per se*, but either (i) an evaluation of a single element in $\bm{K}$, $\bm{P}$, $\bm{G}$, $\bm{Q}$, or $\bm{\hat{Q}}$, or (ii) establishes orders between components of the same type (we do not consider in this paper mixed orders, in which we could determine precedence between some **k** and **g**, but only orders between different **k**, or between different **g**). The difference between these two roles of attitudes stems from the possibility to evaluate individually some component of either $\bm{K}$, $\bm{P}$, $\bm{G}$, $\bm{Q}$, or $\bm{\hat{Q}}$ (as in Ex.9 and Ex.11), and the possibility for that evaluation to amount to a comparison between different components of the same type (as in Ex.10, Ex.12, and Ex.13). We call the first role *optionality* and the second *preference*.

**Optionality.** Favorable or disfavorable evaluation of a single (i.e., independently of others) **k**, **p**, **g**, **q**, or **q̂** is of interest to the software engineer to the extent that it indicates whether it is necessary to define plans that accord with the given **k**, **p**, **g**, **q**, or **q̂**. In other words, the engineer is interested in knowing if some **k**, **p**, **g**, **q**, or **q̂** is *compulsory* or *optional*. It is compulsory if the stakeholders cannot accept a specification that does not account for that particular **k**, **p**, **g**, **q**, or **q̂**; otherwise, it is optional. Consequently, optionality completely partitions each of $\bm{K}$, $\bm{P}$, $\bm{G}$, $\bm{Q}$, and $\bm{\hat{Q}}$ onto optional and compulsory parts: $\bm{K}$ on $\bm{K_O}$ and $\bm{K_C}$, $\bm{G}$ on $\bm{G_O}$ and $\bm{G_C}$, and so on. The dichotomy optional/compulsory arises from the intensity of evaluation: favorable or disfavorable evaluation involves undoubtedly an ill-structured and subjective evaluation space, so that the chosen dichotomy serves to partition that space and thereby simplify the use of evaluations over individual **k**, **p**, **g**, **q**, or **q̂**. The actual intensity becomes of interest once tradeoffs appear. It is then necessary to determine which of two or more optional or compulsory, yet conflicting goals, plans, or otherwise, need to be satisfied. In case of conflict, we speak of preferences and do so regardless of optionality. Note that optionality is a means for comparing alternative specifications, whereby those that satisfy "more" among the optional **k**, **p**, **g**, **q**, or **q̂** is more desired than another that satisfies "less" of these, all other things being equal.

**Preference.** Evaluation that compares two or more instances of the same concept in our ontology (e.g., two different goals) introduces an order between these, and amounts to what we call a preference (order). Ex.10, Ex.12, and Ex.13 establish preference orders as each compares two alternatives in terms of favor or disfavor. Choice of appropriate properties for the model of the preference order (e.g., whether it is transitive) is not of interest in this paper. It is important to note that preferences can be defined over preferences; this is needed when all preferences cannot be simultaneously satisfied. For instance, lower payment verification time may correlate with lower payment security in a flight booking system. When aiming to satisfy one preferred requirement negatively affects the ability to satisfy some other requirement, we say that the involved requirement preferences are conflicting. If the conflict cannot be alleviated through a different system design (e.g., new payment verification servers with better performance both in payment verification and security), the relative importance of the conflicting preferences should be determined. The importance relation between requirement preferences is merely another kind of preference where higher importance corresponds to higher desirability. The following two preference orders over goals give a conflict:

(Ex.14)   *Issuing electronic flight tickets is preferred to issuing paper flight tickets.*

(Ex.15)   *It is preferred that all travelers have paper tickets than only those who have had access to a printer after the booking.*

We must state which of the two preferences is more desirable to follow when choosing among alternative plans. If the preference obtained from the second statement is more important, then a system design which satisfies the first will be less desirable than a system design which satisfies the second but does not satisfy the first. When we say that a preference is satisfied, we mean that the most desirable alternative among the ordered ones is chosen.

## 4   Requirements Problem Revisited

Returning to Zave and Jackson's formulation, if we seek **S** such that for elicited **K** and **R** we have $\mathbf{K}, \mathbf{S} \vdash \mathbf{R}$, we implicitly assume that **K**, **S**, and **R** are precise and complete enough for the satisfaction relation to hold.

We have noted earlier that this is hardly a plausible assumption in practice (§3.4), and we suggested to use a non-monotonic consequence $\mathrel{|\!\sim}$ instead of a monotonic one $\vdash$. This adjustment does not respond to the problem of eliciting and using tentative evaluations: attitudes seem to be irrelevant as they are not covered in $\mathbf{K}, \mathbf{P} \mathrel{|\!\sim} \mathbf{G}, \mathbf{Q}$. They, however, clearly are relevant: the quality of a system is evaluated by the stakeholders over the system's entire lifecycle, leaving the engineer to anticipate during development these future evaluations by eliciting evaluations (i.e., attitudes) and assume that these are representative of the future evaluations.

We have introduced two ways in which attitudes intervene within our ontology, namely through optionality and preference. The former leads us to partition instances of the various ontology concepts into either optional or compulsory, whereas we will denote all preferences by $\mathbf{A}^{\succ}$. Given some preference order $\mathbf{a}^{\succ}$ in $\mathbf{A}^{\succ}$ stating that to satisfy some $\mathbf{g}'$ is preferred to satisfying the alternative $\mathbf{g}$, all other things being equal, we prefer the specification $\mathbf{P}'$, in which plans lead to states described by $\mathbf{g}'$, to the specification $\mathbf{P}$ in which plans lead to states described by $\mathbf{g}$. We can then say that, given $\mathbf{A}^{\succ}$, we seek the most desirable feasible $\mathbf{P}$. Following the definitions outlined above, the most desirable $\mathbf{P}$ is one that brings about the desired and allowed states (that are delimited by requirements and domain assumptions) that rate highest on stakeholders' attitudes. We thus seek $\mathbf{P}$ that satisfies highest ranked requirements and domain assumptions. Since we can have preferences over conflicting preference orders, $\mathbf{P}$ will ideally satisfy highest ranked alternatives that these preferences define. Attitudes that concern the dichotomy optional/compulsory resolve a different problem with Zave and Jackson's formulation: if $\mathbf{K}, \mathbf{S} \vdash \mathbf{R}$ verifies, then all $\mathbf{R}$ obtains. In other words, writing $\mathbf{K}, \mathbf{S} \vdash \mathbf{R}$ makes all requirements compulsory. The optional status is impossible, even though it is clearly encountered in practice.

We can now provide the revised formulation of the requirements problem that accounts for all remarks and the ontology discussed in this paper. Below, $\mathbf{K}$ denotes the finite set of all communicated domain assumptions in a given project. Since we can have alternative domain assumptions (i.e., $\mathbf{k}$ is alternative to $\mathbf{k}'$ if there are no possible states in which both can hold; note that $\mathbf{k}$ and $\mathbf{k}'$ are inconsistent), we can partition $\mathbf{K}$ into a finite number of consistent subsets $\mathbf{K}^i$. Each $\mathbf{K}^i$ therefore contains one alternative for each set of alternatives in $\mathbf{K}$ (i.e., a $\mathbf{K}^i$ contains either $\mathbf{k}$ or $\mathbf{k}'$). We have the same notational convention for $\mathbf{P}$, $\mathbf{G}$, $\mathbf{Q}$, $\hat{\mathbf{Q}}$, and $\mathbf{A}$. The $i$-th consistent subset of the compulsory domain assumptions is thus denoted $\mathbf{K}^i_{\mathbf{C}}$.

**Definition 8.** *Starting from compulsory and optional domain assumptions*

($K_C$ and $K_O$), goals ($G_C$ and $G_O$), quality constraints ($Q_C$ and $Q_O$), softgoals ($\hat{Q}_C$ and $\hat{Q}_O$), plans ($P_C$ and $P_O$), and attitudes ($A_C^\succ$ and $A_O^\succ$) communicated by the stakeholders, the requirements problem amounts to finding the specification (i.e., plans) ($P^*$) such that:

1. $K^*, P^* \mid\!\sim G^*, Q^*, A^\succ$.

2. Preferences in $A^\succ$ indicate that that there is no other combination ($K_C^i, G_C^i, Q_C^i$) that is preferred to ($K_C^*, G_C^*, Q_C^*$).

3. There is no $P^i$ different from $P^*$ such that $K_C^*, P^i \mid\!\sim G_C^*, Q_C^*, A_C^\succ$ verifies, and

   (a) $K_C^*, K_O^i, P^i \mid\!\sim G_C^*, G_O^i, Q_C^*, Q_O^i, A_C^\succ, A_O^\succ$ verifies, where $K_O^i$ is a larger subset of $K_O$ than $K_O^*$, and/or $G_O^i$ is a larger subset of $G_O$ than $G_O^*$, and/or $Q_O^i$ is a larger subset of $Q_O$ than $Q_O^*$, and/or

   (b) Preferences in $A^\succ$ indicate that $K_O^i$ is preferred to $K_O^*$, and/or $G_O^i$ is preferred to $G_O^*$, and/or $Q_O^i$ is preferred to $Q_O^*$.

4. For each softgoal $\hat{q}$ in $\hat{Q}$, there is a quality constraint $q$ that stands in the justified approximation relationship with the softgoal, i.e., $jApprox(\hat{q}, q)$.

5. For each preference order in $A^\succ$ over softgoals, there is a preference order that maintains that same same ordering over quality constraints that stand in the justified approximation relationship to the given softgoals.

Note that the definition assumes (e.g., point 2 in Definition 8) that preferences are combined to obtain an aggregate preference over subsets of $K_C$, $G_C$, and $Q_C$. Observe that since we can define $\mid\!\sim$ in an informal logic (e.g., of argumentation [9]), the above applies if informal notation is used during RE. The engineer no longer seeks a specification which will merely satisfy compulsory requirements, but is expected to define a specification which comes as close as feasible to satisfying all compulsory requirements and as many as feasible optional requirements, and this in the way that rates favorably as high as possible in terms of stakeholders attitudes.

## 5 Conclusions

Given the confusion about the basic ontology in RE, this paper revisits the core concepts and proposes a series of definitions thereof. Definitions are written in a restricted vocabulary of the DOLCE foundational ontology with explicit speech acts and mental states. The scope of the core RE ontology is now determined by the kinds of speech acts that stakeholders communicate to the software engineer. The concepts are shown to cover all speech acts, thereby ensuring that the software engineer can classify all content of stakeholders' communications as instances of concepts or relationships in our ontology. In our core ontology, the concepts are *goal*, *softgoal*, *quality constraint*, *plan*, and *domain assumption*; the relationships are *attitude*-based optionality and preference, *justified approximation*, and *non-monotonic consequence*.

The proposed definitions for goal, quality constraint, and softgoal concepts have several advantages. Together, these concepts cover the key taxonomy of functional and nonfunctional requirements. In contrast to previous work, it is clear which of the nonfunctional requirements are verifiable, and we provided precise conditions for the verifiability of nonfunctional requirements. All three concepts are grounded in an intuitive foundational ontology. It has not been particularly clear how measures fit within the goal and softgoal separation in RE; this issue is resolved herein. While we acknowledge, through quality constraints, the tenet that what cannot be measured cannot be managed, we also make it clear that not all can be directly defined or measured: convenience, security, safety, along with considerations such as fun amount to softgoals, which involve subjective and local quality spaces that are difficult to precisely describe and share. Our definitions are not specific to particular kinds of functional and nonfunctional requirements, but span much of ongoing software engineering research on these topics. We define the softgoal concept in a way that states precisely what a softgoal is within the communication between the engineers and stakeholders, and this without breaking off from the tradition of how softgoals are used in RE (as in [15] and later). We are also in line with research on software measurement and quality: it is indeed now well accepted that considerations such as security, safety, convenience, usability, maintainability, and so on, have no domain- and/or project-independent definition [12, 6]. Our departure from Zave and Jackson lies in the introduction of attitudes and our detailed discussion of functional and nonfunctional requirements. We shown (§4) that Zave and Jackson's **R** only includes our goals, and neither quality constraints nor softgoals. Recently, Glinz suggested that a "nonfunctional requirement is is an attribute of or a constraint on a system" [8], though it remains unclear what concepts the terms 'attribute' and 'constraint' denote. Our conceptualization has the advantages of being grounded in the actual context of RE (i.e., in the communication between the engineer and the stakeholders) and we are explicit on the meaning of our concepts.

The revised ontology leads to a new formulation of the requirements problem, that is, the fundamental problem of the RE field. We have argued that the new formulation fits intuition better and avoids the pitfalls of the previously established formulation from Zave and Jackson [21].

Established RE frameworks lack capabilities needed for the representation of and reasoning about instances of all terms in the terminology. It is for instance unclear how to deal with the justification of approximation in methodological terms, how to elicit and analyze attitudes. These are pressing concern. Appropriate means are necessary to resolve the requirements problem in its revised form.